# EFFECT OF OXYGEN SATURATION ON AFM-WFM-HTSC TRANSITION TEMPERATURES IN $RuSr_2(Eu_{1.5}Ce_{0.5})Cu_2O_{10-\delta}$ CERAMIC SAMPLES

E.Yu. Beliayev, I.G. Mirzoev

*B.I. Verkin Institute for Low Temperature Physics and Engineering, NAS of Ukraine,*

*47 Lenin's Prospect, 61103, Kharkov, Ukraine.*

beliayev@ilt.kharkov.ua

The effect of annealing in high pressure oxygen atmosphere on superconducting transition temperatures for ceramic samples of magnetic superconductor $RuSr_2(Eu_{1.5}Ce_{0.5})Cu_2O_{10-\delta}$ was studied. It was shown that the properties of the samples are consistent with the behavior of granular superconducting system. As a result of oxygen saturation the superconducting transition temperatures become higher. Particularly, the shift of superconducting transition temperature for the intergranular medium is $\Delta T_{ci} = 9.2$ K and for the matter within the granules $\Delta T_{cg} = 6.8$ K. This difference is supposed due to the mechanism of oxygen diffusion along the grain boundaries. In the temperature range $135 < T < 350$ K the behavior of resistance obeyed the Mott's law of variable range hopping for three-dimensional case.

## 1. Introduction

Continuously improving technology of creating complex oxide compounds aimed at the creation of materials that demonstrate a combination of properties that previously considered being impossible. One such material is europium and gadolinium ruthenocuprates family that combine the properties of superconducting and the magnetically ordered states – the so-called "magnetic superconductor". In ruthenocuprates $RuSr_2(Eu_{1.5}Ce_{0.5})Cu_2O_{10-\delta}$ two competing order parameters – the spontaneous magnetization, with emerging at $T_M$ (100 ÷ 140 K) weak ferromagnetic component in the sublattice of Ru atoms, and superconducting order parameter due to hole doping of $CuO_2$ layers with critical temperature $T_c$ (30 ÷ 50 K), – coexist in one crystal lattice, experiencing mutual spatial modulation.

The possibility of coexistence of superconducting and magnetic properties in these materials is due to the tilting of antiferromagnetically ordered magnetic moments in the Ru sublattice [1] that induces on $CuO_2$ conducting layers inner magnetic field of $H \approx 0.1$ T [2]. The value of this magnetic field is greater than the lower critical field $H_{C1}$ (estimated $\approx 5 \times 10^{-3}$ T [3]), but significantly less than the upper critical field $H_{C2} \approx 28 \div 80$ T [4]. Thus, the superconductivity in $CuO_2$ layers germinates directly in the mixed state [2].

However, being, in fact, high-temperature superconductors (HTSC) based on copper oxide layers, ruthenocuprates inherit a tendency of copper oxide high-temperature superconductors to lose the stoichiometric oxygen included in their crystal lattice, which leads to degradation of their superconducting properties.

The aim of this work was to study the effect of annealing in an atmosphere of pure oxygen at high pressure on the characteristic temperatures of inter-granular and intra-granular superconducting transitions for granular ceramic samples of magnetic superconductor $RuSr_2(Eu_{1.5}Ce_{0.5})Cu_2O_{10-\delta}$.

**2. Experimental procedure**

Test samples were prepared in the «Rakah Institute of Physics» by solid-phase synthesis. One of the samples obtained was left in the original (as prepared) state (Sample Eu_A). Two others were annealed for 24 hours at the temperature $T = 900°$ C and high pressure of pure oxygen $P = 100$ atm. (Sample Eu_B), and $P = 50$ atm. (Sample Eu_C) [5]. Powder diffraction studies confirmed the purity and prevailing single-phase composition (~ 97 %) for the samples obtained. The unit cell parameters ($a = b = 3.846$ Å, $c = 28.72$ Å) within the experimental error coincided with the data for europium ruthenocuprates given in the works of other authors [3].

For electrical measurements the pressed ceramic tablets were cut in pieces having the form of parallelepipeds with dimensions $10 \times 2 \times 1$ mm$^3$ and distance between the potential contacts was 4 mm.

The measurements of the temperature dependences of resistance were carried out in Kharkov Institute for Low Temperature Physics and Engineering (ILTPE) in the temperature range $3.5 \div 350$ K by 4-probe DC technique applying the current $I = 100$ mkA. The samples were mounted on a copper sample holder in a vacuum chamber of the cryostat. To improve the thermal contact the heat-conducting glue was used. The temperature in the cryostat was controlled by automatic stabilization system with accuracy from $\approx 10^{-4}$ K in the region of liquid helium temperatures to $\approx 10^{-1}$ K at the maximum achievable temperature of 350 K. In the experimental setup we used a stabilized DC power supply with alternating polarity to compensate for the thermal EMF, graded Pt and $RuO_2$ resistive thermometers, multimeters Keithley 2000 and nanovoltmeter Keithley 2182.

### 3. Results and discussion

The temperature dependences of resistance for all three samples studied are shown in Fig. 1. The granular structure of ceramic samples manifested itself in the typical "shouldered" form of the observed intra-granular and inter-granular resistive transitions [6].

From Fig. 1 we can see that for the samples saturated with oxygen almost twofold decrease in resistance was observed in the normal state, apparently due to the injection of additional current carriers – holes – into the layers of $CuO_2$.

Fig. 2 shows the first derivatives for the temperature dependences of resistance, which allow determining, by positions of the peaks, the critical temperatures corresponding to superconducting transition both for the substance within the granules and for the intergranular environment in the granular systems under study.

Due to the fact that the ceramic samples studied were granular systems, the transition to the superconducting state occurred in a two step process. For $T < 50$ K at first we see the establishing of superconducting state for the substance within the granules. The critical temperatures of that transition varied in the range from $T_{cg} = 38.6$ to 45.4 K depending on the degree of oxygen saturation. Then there follows a small

"shoulder", after which we can see a further drop in resistance associated with the establishment of Josephson weak links in the disordered intergranular medium, with critical temperatures that vary from $T_{ci}$ = 10 K for the "as prepared" sample Eu_A to $T_{ci}$ = 19.2 K for the samples Eu_B and Eu_C annealed in oxygen at different pressures (See Fig. 2).

As a result of oxygen saturation (Fig. 2) we can determine the displacement of superconducting transition temperatures for intergranular medium $\Delta T_{ci}$ = 9.2 K and for the matter within the granules $\Delta T_{cg}$ = 6.8 K. This difference is obviously due to the mechanism of oxygen diffusion along the grain boundaries. The rate of oxygen diffusion is only slightly dependent on pressure, so resistance curves for the samples annealed in oxygen at pressures of 50 and 100 atm are practically identical. The large width of superconducting transition for intergranular environment (with corresponding $d\rho/dT$ peak half-width ≈ 18 K in Fig. 2) can be explained by its strong heterogeneity. After annealing in oxygen, judging by the increase in the width of intergranular superconducting transition, this heterogeneity was even intensified, which may be the evidence of further decomposition of intergranular medium caused by partial evaporation of the most volatile component of compound – RuO$_2$ [7].

For all three of the samples studied in high-temperature range ($T$ = 135 ÷ 350 K) the temperature dependence of resistance obeyed the Mott's hopping conductivity law, inherent to the mechanism of variable range hopping (VRH) for three-dimensional case,

$$\rho \approx \rho_0 \exp(T_0/T)^{1/4}, \qquad (1)$$

where $T_0 = B_0^4 \cdot (L_c^{-3}/k_B \cdot N(E_F))$ – Mott's characteristic temperature (in fact it is a total sum of energy of all the charge carriers in the mean localized volume, expressed in the temperature units), $B_0$ = const = 1.7 ÷ 2.5, $k_B$ – Boltzmann constant. This conductivity behavior corresponded to hopping conductivity between the isolated granules of ruthenocuprate which were at these temperatures in a non-superconducting state (see Fig. 3). When fitting the experimental data we obtained the values $T_0$ ≈ 150,000 K for Eu_A and $T_0 \approx 75{,}000$ K for samples Eu_B and Eu_C, which are common values if you compare them with the values of Mott's temperature obtained by other authors [8].

The lack of data on the density of electronic states at the Fermi level $N(E_F)$ for, still little-studied, europium-based ruthenocuprates does not allow to calculate the radii of localization of the wave functions of charge carriers $L_c$.

For temperatures $T < 135$ K resistance of ruthenocuprate samples decreases with temperature more rapidly than predicted by Mott's law. The reason for this drop of resistance may be a ferromagnetic ordering in the samples studied. Marked by an arrow in Fig. 3 the temperature $T = 135$ K, corresponds to the temperature of kink on the temperature dependence of magnetization measured in ZFC (zero field cooling) regime for sample Eu_C [5]. After this temperature magnetization of the sample increases and a feature on the temperature dependence of thermal conductivity was observed [5] which may also be connected with some processes of structural relaxation during AFM-WFM magnetic transition.

It should be noted that the mere appearance of a weak ferromagnetic state in ruthenocuprates is caused by the fact that the covalent bonds lengths for Cu-O and Ru-O, being almost equal at room temperature, have different temperature dependence. With lowering temperature, the Cu-O bonds shrink more than Ru-O. This leads to coordinated rotation of adjacent oxygen octahedra which become too big to fit the lattice and in which the antiferromagnetically ordered $Ru^{+5}$ ions are confined [9]. This rotation achieves the angle $14°$ and through Dzyaloshinskii-Moriya interaction causes the coordinated tilting of antiferromagnetically ordered $Ru^{+5}$ magnetic moments perpendicular to the layered structure of the compound. At $T_m \approx 135$ K in this system of uncompensated perpendicular component of antiferromagnetically ordered magnetic moments of $Ru^{+5}$ ions even small magnetic field (from Earth or cryostat) causes weak ferromagnetic (WFM) (actually component ferrimagnetic) state. This ferromagnetic ordering in perpendicular components of magnetization results in a reduction of resistance by reducing the magnetic scattering of the conduction electrons. Apparently, this process of establishing of ferromagnetic order in ruthenocuprate samples determines the temperature behavior of resistance in the temperature range between the $T_m$ and $T_{cg}$ and explains the violation of the initial VRH Mott's low.

The influence of an external magnetic field on superconducting transition in granulated ruthenocuprates is shown by the example of sample Eu_A (as prepared) in Fig. 4. The Figure shows that application of weak magnetic field $H = 0.01$ T doesn't change the superconducting transition within the granules because the upper critical field in ruthenocuprates is very high ($H_{C2} \approx 28 \div 80$ T [4]) and thanks to the ferromagnetic component of the internal magnetic field the superconducting material in granules is already in a mixed state. But this small magnetic field easily destroys a weak superconductivity in disordered intergranular environment.

## 4. Conclusions

Thus, by studying the temperature and magnetic field dependences of resistance for ceramic samples of europium ruthenocuprate $RuSr_2(Eu_{1.5}Ce_{0.5})Cu_2O_{10-\delta}$ we investigate the influence of annealing in high pressure oxygen atmosphere on superconducting transition temperatures both inside granules and in disordered intergranule media. The investigated samples showed the temperature and magnetic field behavior pertinent to granular superconducting system. It was shown that oxygenation rises the superconducting transition temperatures both inside the granules and in the disordered intergranular environment. A stronger shift of the superconducting transition temperature for intergranular medium compared with the intragranular one indicates that the penetration of oxygen into the sample during annealing occurs along the grain boundaries. The rate of oxygen diffusion is only slightly dependent on pressure. It was shown that the temperature behavior of the samples' resistance above the temperature of magnetic transition $T_m = 135$ K matches the Mott's formula for variable range hopping in three-dimensional case. However, in the intermediate temperature range ($T_{cg} < T < T_m$), when the ferromagnetic order in the sample had already been established, and superconductivity has not yet occurred, there are deviations from the Mott's law, apparently related to the fact that the appearance of a weak ferromagnetic order and structural relaxation in crystal lattice lead to reduction in charge carriers scattering.

The author is grateful to his teacher Prof. B.I. Belevtsev and his colleague Prof. D.G. Naugle from Texas A&M University for the samples provided for these measurements.

**Figure captions**

Fig. 1. Temperature dependences of resistance for three samples of RuSr$_2$(Eu$_{1.5}$Ce$_{0.5}$)Cu$_2$O$_{10-\delta}$.

Fig. 2. Influence of annealing in oxygen on superconducting transition temperatures.

Fig. 3. VRH mechanism for temperatures $T > T_m$ in antiferromagnetic (AFM) state of ruthenocuprate samples Eu_A, Eu_B and Eu_C. For temperatures $T_{cg} < T < T_m$ after transition to weak ferromagnetic (WFM) state deviations from Mott's law arise.

Fig. 4. Effect of a weak magnetic field on superconductivity in intergranular medium.

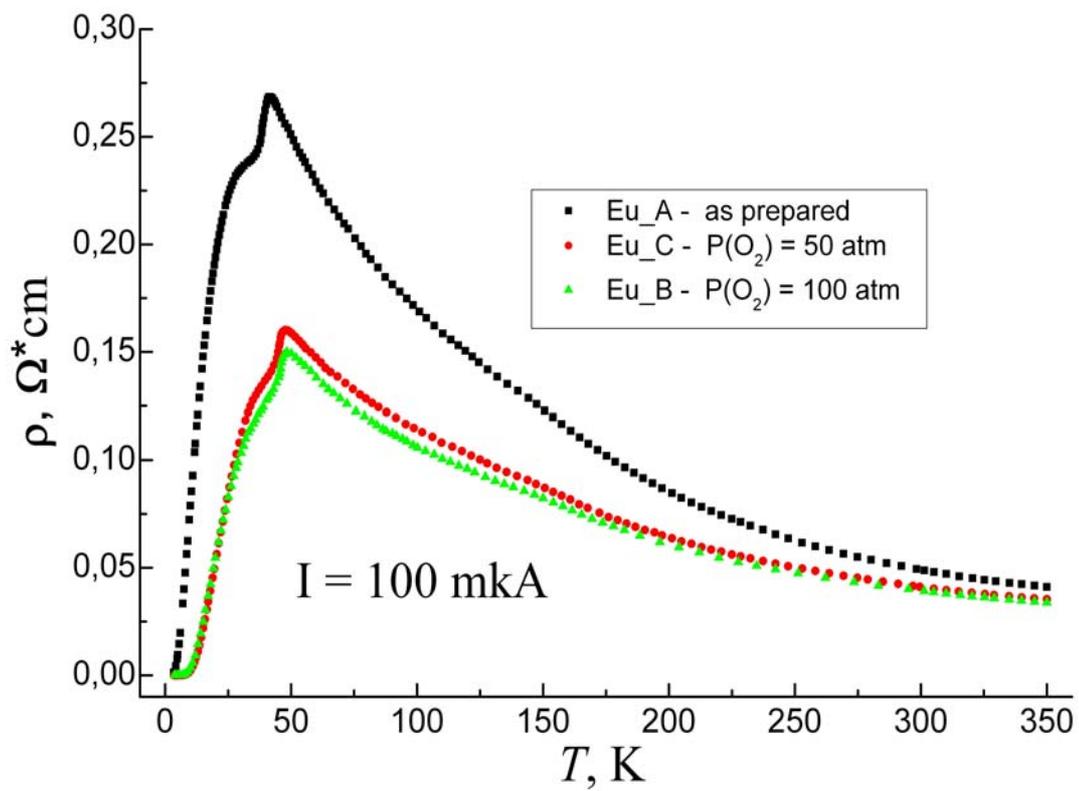

**Fig. 1**
E.Yu. Beliayev, I.G. Mirzoev
**Effect of Oxygen Saturation on AFM-WFM-HTSC Transition Temperatures
in RuSr$_2$(Eu$_{1.5}$Ce$_{0.5}$)Cu$_2$O$_{10-\delta}$ Ceramic Samples**

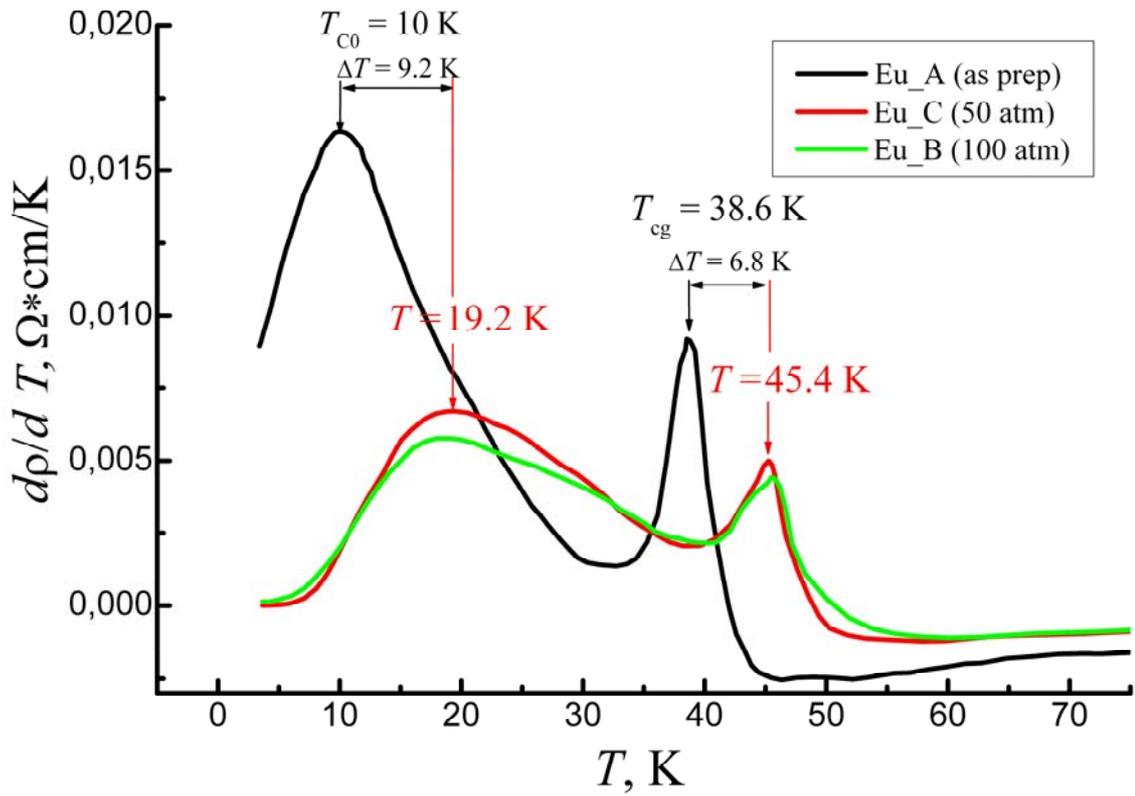

**Fig. 2**
E.Yu. Beliayev, I.G. Mirzoev
**Effect of Oxygen Saturation on AFM-WFM-HTSC Transition Temperatures in $RuSr_2(Eu_{1.5}Ce_{0.5})Cu_2O_{10-\delta}$ Ceramic Samples**

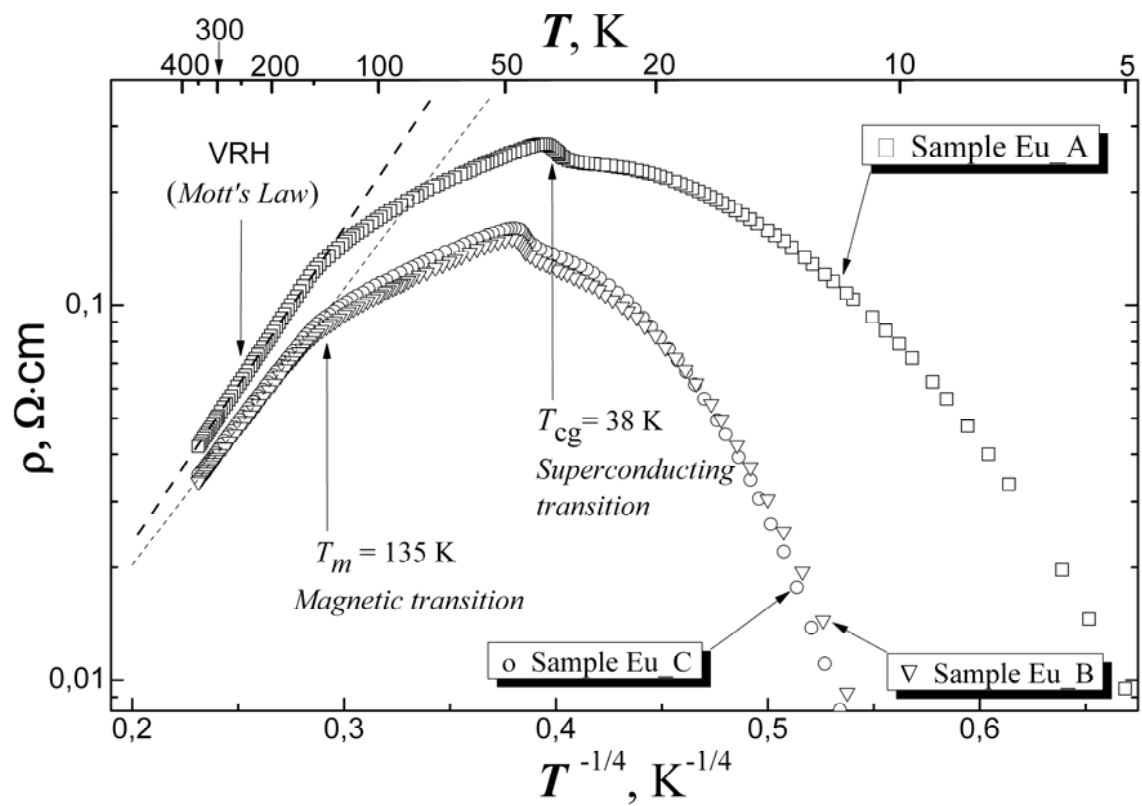

**Fig. 3**
E.Yu.Beliayev.
E.Yu. Beliayev, I.G. Mirzoev
**Effect of Oxygen Saturation on AFM-WFM-HTSC Transition Temperatures in $RuSr_2(Eu_{1.5}Ce_{0.5})Cu_2O_{10-\delta}$ Ceramic Samples**

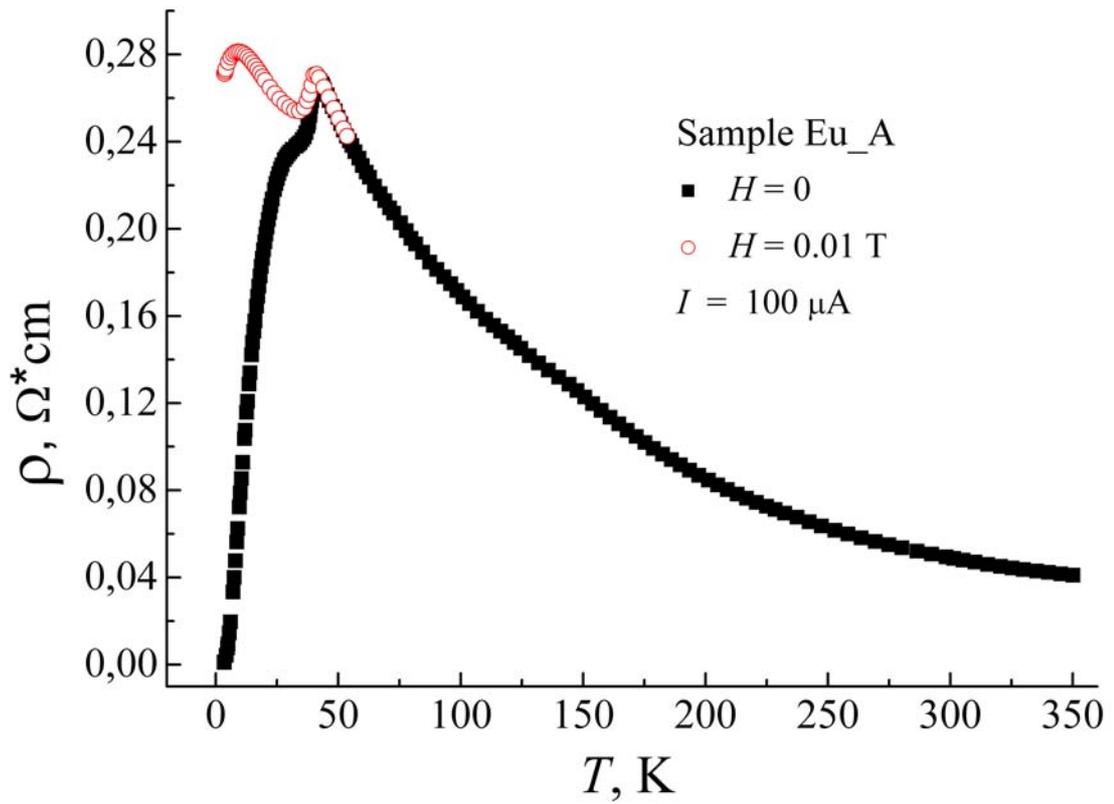

**Fig. 4**
E.Yu. Beliayev, I.G. Mirzoev
**Effect of Oxygen Saturation on AFM-WFM-HTSC Transition Temperatures in RuSr$_2$(Eu$_{1.5}$Ce$_{0.5}$)Cu$_2$O$_{10-\delta}$ Ceramic Samples**